\documentclass[conference,onecolumn,10pt]{IEEEtran}
\usepackage{enumerate}
\usepackage{eurosym}
\usepackage{amsmath}
\usepackage{amsfonts}
\usepackage{amssymb}
\usepackage{graphicx}
\usepackage{epstopdf}
\usepackage{setspace}
\usepackage{arydshln}
\usepackage{float}
\usepackage{algorithm}
\usepackage[noend]{algpseudocode}
\usepackage[english]{babel}

\begin{document}

\title{End-to-end Learning of Waveform Generation and Detection for Radar Systems}
\author{\IEEEauthorblockN{Wei Jiang\IEEEauthorrefmark{1}, Alexander M. Haimovich\IEEEauthorrefmark{1}, and Osvaldo Simeone\IEEEauthorrefmark{1}\IEEEauthorrefmark{2}}\\
	\IEEEauthorblockA{\IEEEauthorrefmark{1}CWiP, New Jersey Institute of Technology, Newark, New Jersey 07102, USA \\
		\IEEEauthorrefmark{2}KCLIP lab, Department of Engineering, King's College London, London, WC2R 2LS, UK
	\\Email: {\{wj34, haimovic\}@njit.edu}, osvaldo.simeone@kcl.ac.uk} }
\maketitle

\begin{abstract}
An end-to-end learning approach is proposed for the joint design of transmitted waveform and detector in a radar system. Detector and transmitted waveform are trained alternately: For a fixed transmitted waveform, the detector is trained using supervised learning so as to approximate the Neyman-Pearson detector; and for a fixed detector, the transmitted waveform is trained using reinforcement learning based on feedback from the receiver. No prior knowledge is assumed about the target and clutter models. Both transmitter and receiver are 
implemented as feedforward neural networks. Numerical results show that the proposed end-to-end learning approach is able to obtain a more robust radar performance in clutter and colored noise of arbitrary probability density functions as compared to conventional methods, and to successfully adapt the transmitted waveform to environmental conditions.
\end{abstract}



\section{Introduction}

Optimal waveform design and target detection have long been topics of central interest in radar \cite{Kay1998}, \cite{Kay 2007}. The traditional design of optimal radar detectors and
optimal radar waveforms relies on the mathematical modeling of many environmental aspects, including the statistics of targets, clutter, and noise. When the mathematical models are complex, optimal solutions may not be available, or they may be too computationally intensive to implement \cite{Farina 1987}, \cite{Gini 2012}.  Moreover, optimized solutions are generally not robust when the actual statistics of the environment deviate from the assumed models.


Machine learning has been successfully applied to solve problems for which reliable mathematical models are unavailable or too complex to yield feasible optimal solutions, such as in computer vision and natural language processing. Recently, machine learning has been applied to the design of the physical layer in communication systems. Notably, in \cite{OShea 2017}, it is proposed to jointly train encoder and decoder of a communication link by treating the cascade of encoder, channel, and decoder as an autocoder \cite{deeplearning}, \cite{osvaldo1}. This approach requires the availability of a known channel model. For the case of an unknown channel model, reference \cite{Aoudia 2018} proposes an alternate training approach, whereby the transmitter is trained via reinforcement learning through feedback from the receiver, while the decoder is trained using supervised learning. In order to enable reinforcement learning, a loss metric is measured at the receiver and communicated over a reliable channel to the transmitter. A detailed review of the state of the art can be found in \cite{osvaldo2} (see also \cite{osvaldo3} for recent work).


In the radar field, machine learning-based approaches have been suggested for implementing Neyman-Pearson (NP) detectors in \cite{Moya 2009}, \cite{Moya 2013}. In these papers, learning machines trained in a supervised manner using a suitable loss function are proved to approximate the NP detector. As a representative example, in \cite{Moya 2013}, a neural network is trained to implement a radar detector assuming unknown statistical models for the radar channel using supervised learning. In such case, a conventional NP detector is
intractable, since the likelihood ratio cannot be computed. The authors show
that the performance of the neural network detector is comparable with that
of the NP detector obtained in the ideal case in which the model is known.


In this work, we introduce an end-to-end learning approach for the joint design of waveform and detector in a radar system. Unlike the traditional design of radar systems that assumes knowledge of mathematical models for target, clutter, and noise, the proposed learning-based design relies on data, and is hence able to adapt to the actual statistics of the environment. Inspired by \cite{Aoudia 2018}, an alternate learning procedure is introduced whereby the receiver is trained via supervised learning while the transmitted waveform is held fixed; and the design of the waveform is carried out via reinforcement learning for a fixed receiver design. Learning of the transmitter and receiver are alternated until a stopping criterion is satisfied.


\section{System description}
\begin{figure}
	\vspace{-41ex}
	\hspace{2ex}
	\includegraphics[width=1.3\linewidth]{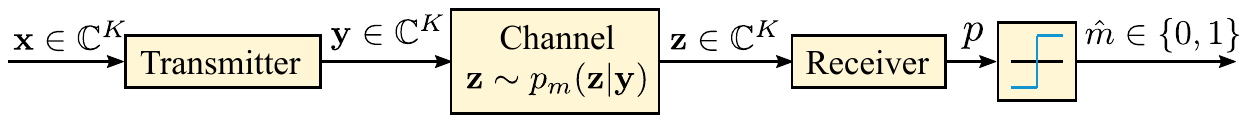}
	\vspace{-160ex}
	\caption{End-to-end learning of a radar system.}
	\vspace{-3ex}
	\label{f:end_to_end}
\end{figure}

We focus on the radar system illustrated in Fig. \ref{f:end_to_end}, in which a transmitter and a receiver form a system seeking to detect the presence of a target. Both the transmitter and the receiver are implemented as two separate parametric functions $f_{\boldsymbol{\theta}_T}(\cdot)$ and $f_{\boldsymbol{\theta}_R}(\cdot)$ with trainable parameter sets $\boldsymbol{\theta}_T$ and $\boldsymbol{\theta}_R$, respectively. As shown in Fig. \ref{f:end_to_end}, the input to the transmitter is a user-defined initialization waveform $\mathbf{x} \in \mathbb{C}^K$ containing $K$ complex chips. The output of the transmitter is the radar waveform $\mathbf{y}\in \mathbb{C}^{K}$ obtained through a trainable mapping $\mathbf{y}=f_{\boldsymbol{\theta}_T}(\mathbf{x})$. The radar waveform $\mathbf{y}$ is transmitted via a radar channel $ p_m(\mathbf{z}|\mathbf{y})$, which acts as a stochastic system, to produce the observation vector $\mathbf{z}\in \mathbb{C}^K$ of the detector. The channel depends on the presence or absence of a target, which is represented by the binary variable $m$, taking $m=1$ and $m=0$ according to the presence or absence of target. The receiver passes the observation vector $\mathbf{z}$ through a trainable mapping $p=f_{\boldsymbol{\theta}_R}(\mathbf{z})$, which produces the scalar $p\in (0,1)$. The final decision $\hat{m}\in\{0,1\}$ is made by comparing the output of the receiver $p$ to a hard threshold in the interval $(0,1)$.

We aim to jointly optimize trainable parameter sets $\boldsymbol{\theta}_T$ and $\boldsymbol{\theta}_R$ of two functions implementing transmitter $f_{\boldsymbol{\theta}_T}(\cdot)$ and receiver $f_{\boldsymbol{\theta}_R}(\cdot)$ to meet application-specific performance requirements. The training process consists of receiver training iterations and transmitter training iterations. A receiver training iteration optimizes the receiver parameter set $\boldsymbol{\theta}_R$ for a fixed transmitter parameter set $\boldsymbol{\theta }_{T}$, while a transmitter training iteration optimizes $\boldsymbol{\theta }_{T}$ for a fixed receiver parameter set $\boldsymbol{\theta }_{R}$. Details of training and optimization of the receiver and transmitter are presented next. The joint training procedure of the receiver and transmitter is summarized for reference in Algorithm 1.

As mentioned in the Section I, the end-to-end learning approach does not have to rely on rigid mathematical models. Thus, if available, real data can be applied to train the system. Alternatively, if real data is not available, synthetic data may be used for training. Reliance on rigid mathematical models may be avoided even in this instance. For example, in traditional radar design, an optimal detector is derived based on an assumed model for target and interference. However, as will be demonstrated by numerical results, the synthetic data based used for training the end-to-end radar system may be generated by multiple models.

\subsection{Receiver Training}

For a given transmitted waveform $\mathbf{y}$, a training sample vector $\mathbf{z}$ is generated according to the channel model $p_m(\mathbf{z}|\mathbf{y})$.
The cross-entropy \cite{Moya 2013} is adopted as the loss function for the receiver
\begin{equation}
\begin{aligned}
	L_R(\boldsymbol{\theta}_R)=-\mathbb{E}_{ \substack{ m\sim p(m),\\ \mathbf{z}\sim p_m(\mathbf{z}|\mathbf{y}) } } \bigg\{ m \log f_{\boldsymbol{\theta}_R}(\mathbf{z})+ (1-m)\log\big[1-f_{\boldsymbol{\theta}_R}(\mathbf{z}) ]  \bigg\}. \label{eq:CE rx loss}
\end{aligned}
\end{equation}
The ensemble loss (\ref{eq:CE rx loss}) requires averaging over the distribution of the target presence distribution $m$ and of the radar channel $p_m(\mathbf{z}|\mathbf{y})$, which are both unknown. To tackle this problem, we assume the availability of $Q_R$ independent samples drawn from these distributions, and we let $(m_q, \mathbf{z}_q)\sim p(m)p_m(\mathbf{z}|\mathbf{y}=f_{\boldsymbol{\theta}_T}(\mathbf{x}))$ represent the $q$th training sample. Since $p(m)$ is not known, it is assumed $p(m)=1/2$. The ensemble loss (\ref{eq:CE rx loss}) is then estimated by the training cross-entropy loss
\begin{equation}
\begin{aligned}
	\hat{L}_R(\boldsymbol{\theta}_R)=-\frac{1}{Q_R}\sum_{q=1}^{Q_R}\bigg\{ m_q \log f_{\boldsymbol{\theta}_R}(\mathbf{z}_q) + (1-m_q)\log\big[1-f_{\boldsymbol{\theta}_R}(\mathbf{z}_q) ]  \bigg\}. \label{eq: est CE rx loss}
\end{aligned}
\end{equation}
From (\ref{eq: est CE rx loss}), the receiver parameter set $\boldsymbol{\theta}_R$ is trained for a fixed transmitted waveform, specified by vector $\boldsymbol{\theta}_T$, by tackling the problem
\begin{equation}
\boldsymbol{\theta }_{R}^{\ast }=\arg \min_{\boldsymbol{\theta }_{R}} \hat{L}_R(\boldsymbol{\theta}_R). \label{eq: min_R}
\end{equation}

\begin{figure}[H]
	\vspace{-2ex}
    \hspace{8ex}
	\includegraphics[width=1.3\linewidth]{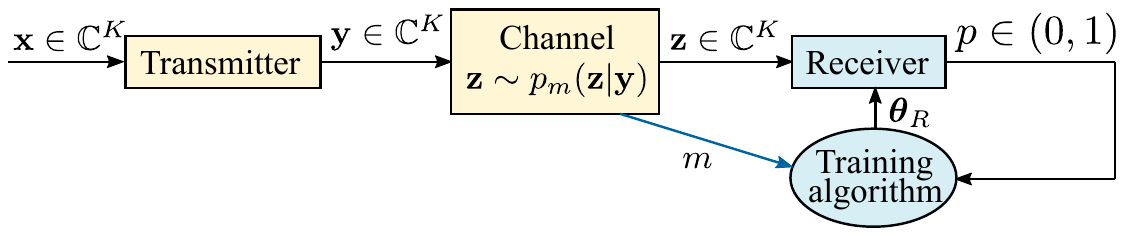}
	\vspace{-195ex}
	\caption{Supervised training of the receiver.}
	\vspace{-2ex}
	\label{f:ml-rx}
\end{figure}

Assuming that the function $f_{\boldsymbol{\theta }_R}(\cdot)$ implementing the receiver is differentiable with respect to the trainable parameter set $\boldsymbol{\theta}_R$, as is the case for feedforward neural networks discussed in Section III, the stochastic gradient descent (SGD) algorithm, or one of its variants \cite{SGD}, can be applied to perform the optimization in (\ref{eq: min_R}). We use superscript $j$ to identify the iteration of the SGD algorithm.  At the $j$th iteration, the receiver parameter set $\boldsymbol{\theta}_R$ is updated according to the SGD rule
\begin{equation}
\boldsymbol{\theta }_{R}^{\left( j+1\right) }=\boldsymbol{\theta }%
_{R}^{\left( j\right) }-\eta {\nabla }_{\boldsymbol{\theta }_{R}}\hat{L}_R(\boldsymbol{\theta}_R^{(j)}),  \label{eq: rx sgd}
\end{equation}
where ${\nabla }_{\boldsymbol{\theta }_{R}}\hat{L}_R(\boldsymbol{\theta}_R^{(j)})$ is the gradient of the training loss $\hat{L}_R(\boldsymbol{\theta}_R)$ with respect to the receiver parameter set $\boldsymbol{\theta}_R$ evaluated at $\boldsymbol{\theta}_R=\boldsymbol{\theta}_R^{(j)}$, and $\eta >0$ is the learning rate. The supervised training of the receiver for fixed transmitter's parameters $\boldsymbol{\theta}_T$ is illustrated in Fig. \ref{f:ml-rx}. 

\subsection{Transmitter Training}
In the transmitter training stage, the receiver parameter set $\boldsymbol{\theta }_R$ is kept fixed, and the function $f_{\boldsymbol{\theta}_T}(\cdot)$ implementing the transmitter is optimized. The goal of transmitter training is to find an optimized parameter vector $\boldsymbol{\theta}_T$ that ideally minimizes the cross-entropy loss in (\ref{eq:CE rx loss}). Following \cite{Aoudia 2018}, we formulate the problem by introducing a randomized transmitter's policy that outputs a waveform $\mathbf{y}$ with probability $\pi_{\boldsymbol{\theta}_T}(\mathbf{y}|\mathbf{x})=\mathcal{N}(\mathbf{y}|f_{\boldsymbol{\theta}_T}(\mathbf{x}),\sigma^2)$, where $\mathcal{N}(\cdot|\mu,\sigma^2)$ represents the Gaussian distribution with mean $\mu$ and variance $\sigma^2$. Randomization is useful to enable exploration of the space of transmitted waveforms and to simplify the gradient-based minimization of the cross-entropy loss with respect to $\boldsymbol{\theta}_T$.

\begin{figure}[H]
	\vspace{-2ex}
	\hspace{6ex}
	\includegraphics[width=1.3\linewidth]{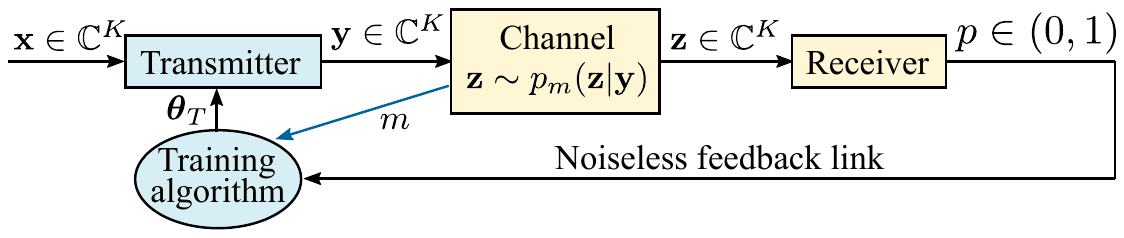}
	\vspace{-196ex}
	\caption{Reinforcement learning-based transmitter training.}
	\vspace{-2ex}
	\label{f:ml-trans}
\end{figure}

To elaborate, for the training of the transmitter, we aim at minimizing the ensemble cross-entropy loss
\begin{equation}
\begin{aligned}
L_T(\boldsymbol{\theta}_T)=-\mathbb{E}_{ \substack{ m\sim p(m),\\ \mathbf{y}\sim \pi_{\boldsymbol{\theta}_T}(\mathbf{y}|\mathbf{x}),\\ \mathbf{z}\sim p_m(\mathbf{z}|\mathbf{y}) }   }  \bigg\{  m \log f_{\boldsymbol{\theta}_R}(\mathbf{z})
+ (1-m)\log\big[1-f_{\boldsymbol{\theta}_R}(\mathbf{z}) ]  \bigg\}. \label{eq:CE tx loss}
\end{aligned}
\end{equation}
Unlike the receiver-side loss (\ref{eq:CE rx loss}), it is not possible to directly approximate (\ref{eq:CE tx loss}) using samples from the distribution $p(m)\pi_{\boldsymbol{\theta}_T}(\mathbf{y}|\mathbf{x})p_m(\mathbf{z}|\mathbf{y})$, since the latter depends on the parameters $\boldsymbol{\theta}_T$ under optimization. To deal with this problem, we leverage the policy gradient theorem \cite{Sutton 2000}, which states that the gradient of the ensemble loss (\ref{eq:CE tx loss}) can be written as
\begin{equation}
	{\nabla }_{\boldsymbol{\theta }_{T}} {L}_T(\boldsymbol{\theta}_T)=\mathbb{E}_{\substack{ m\sim p(m),\\ \mathbf{y}\sim \pi_{\boldsymbol{\theta}_T}(\mathbf{y}|\mathbf{x}),\\ \mathbf{z}\sim p_m(\mathbf{z}|\mathbf{y}) }} \bigg[ l(m,\mathbf{z}) {\nabla }_{\boldsymbol{\theta }_{T}}\log \pi_{\boldsymbol{\theta}_T}(\mathbf{y}|\mathbf{x})  \bigg], \label{eq:gradient tx loss}
\end{equation}
where $l(m,\mathbf{z})=-\big\{  m \log f_{\boldsymbol{\theta}_R}(\mathbf{z})  + (1-m)\log\big[1-f_{\boldsymbol{\theta}_R}(\mathbf{z}) ]  \big\}$ can be interpreted as the instantaneous value of the loss for a pair $(m,\mathbf{z})$. The gradient (\ref{eq:gradient tx loss}) can be estimated via $Q_T$ samples $(m_q,\mathbf{y}_q,\mathbf{z}_q)$ drawn independent and identically distributed (i.i.d.) from the distribution $p(m)\pi_{\boldsymbol{\theta}_T}(\mathbf{y}|\mathbf{x})p_m(\mathbf{z}|\mathbf{y})$, yielding
\begin{equation}
\nabla _{\boldsymbol{\theta }_{T}} \hat{L}_T(\boldsymbol{\theta}_T)=\frac{
	1}{Q_T}\sum_{q=1}^{Q_T}\bigg[ l(m_q,\mathbf{z}_q) \nabla _{\boldsymbol{\theta }_{T}}\log \pi_{\boldsymbol{\theta}_T}(\mathbf{y}_q|\mathbf{x}) \bigg].  \label{eq:tx loss grad}
\end{equation}
At the $j$th iteration of the SGD
algorithm, the transmitter parameter set $\boldsymbol{\theta }_T$ is updated according to the stochastic gradient update rule
\begin{equation}
\boldsymbol{\theta }_{T}^{\left( j+1\right) }=\boldsymbol{\theta }%
_{T}^{\left( j\right) }-\eta {\nabla }_{\boldsymbol{\theta }_{T}} \hat{L}_T(\boldsymbol{\theta}_T^{(j)}).  \label{eq: tx sgd}
\end{equation}
To make this possible, the instantaneous losses $l(m_q,\mathbf{z}_q)$ for $q=1,\dots,Q_T$ are computed by the receiver and then forwarded to the transmitter via a noiseless feedback channel. The reinforcement-learning based training of the transmitter for fixed receiver's parameters $\boldsymbol{\theta}_R$ is illustrated in Fig. \ref{f:ml-trans}.

\begin{algorithm}[H]
	\caption{End-to-End Training of a Radar System}
	\begin{algorithmic}
		\State {\textbf{Step 0}}: Select initialization waveform $\mathbf{x}$, initialize $\boldsymbol{\theta}_R^{(0)}$, $\boldsymbol{\theta}_T^{(0)}$, and set $j=0$
		
		\State {\textbf{Step 1}}: Find $\boldsymbol{\theta}_R^{(j)}$ by minimizing the estimated loss $L_R(\boldsymbol{\theta}_R)$ (\ref{eq:CE rx loss})  with $\boldsymbol{\theta}_T=\boldsymbol{\theta}_T^{(j-1)}$ (see Section II-A)
		
		\State {\textbf{Step 2}}: Find $\boldsymbol{\theta}_T^{(j)}$ by minimizing the estimated loss $L_T(\boldsymbol{\theta}_T)$ (\ref{eq:CE tx loss}) with $\boldsymbol{\theta}_R=\boldsymbol{\theta}_R^{(j)}$ (see Section II-B)
		
		\State {\textbf{Step 3}}: Set $j = j+1$
		\State {\textbf{Step 4}}: Repeat steps 1 to 3 until a stopping criterion is satisfied
	\end{algorithmic}
\end{algorithm}

\section{generic Transmitter and Receiver Architectures}
The algorithm developed in Section II can be applied to any pairs of differentiable functions implementing the transmitter $f_{\boldsymbol{\theta}_T}(\cdot)$ and the receiver $f_{\boldsymbol{\theta}_R}(\cdot)$. In this section, we detail an implementation based on feedforward neural networks. 

A feedforward neural network is a parametric function $f_{\boldsymbol{\theta}}(\cdot)$ that maps an input real-valued vector $\mathbf{r}_0\in \mathbb{R}^{N_0}$ to an output real-valued vector $\mathbf{r}_I \in \mathbb{R}^{N_I}$ via $I$ successive layers. At the output of the $i$th layer, the intermediate output is given by
\begin{equation}
	\mathbf{r}_i=f_{\boldsymbol{\theta}^{[i]}}(\mathbf{r}_{i-1})=\phi\big( \mathbf{W}^{[i]}\mathbf{r}_{i-1} +\mathbf{b}^{[i]}  \big), \text{ } \text{for}\text{ } i=1,\dots,I,
\end{equation}
where $\boldsymbol{\theta}^{[i]}=\{ \mathbf{W}^{[i]}, \mathbf{b}^{[i]} \}$ is the trainable parameter set for the $i$th layer, which includes the weight $\mathbf{W}^{[i]}$ and the bias $\mathbf{b}^{[i]}$, and $\phi(\cdot)$ is an activation function. The set of trainable parameters of the neural network consists of all layers' parameters $\boldsymbol{\theta}=\{\boldsymbol{\theta}^{[1]}, \cdots, \boldsymbol{\theta}^{[I]}\}$.

\begin{figure}[H]
	\vspace{-3ex}
	\hspace{14ex}
	\includegraphics[width=1.3\linewidth]{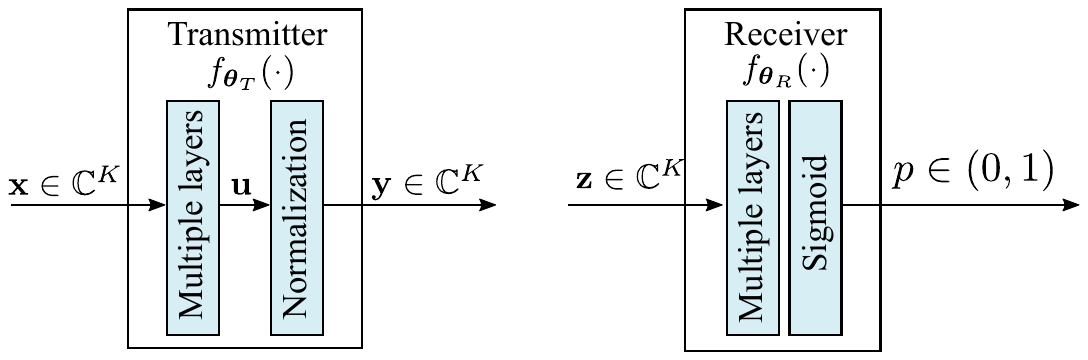}
	\vspace{-185ex}
	\caption{Transmitter and receiver architectures based on feedforward neural networks.}
	\label{f: arch}
\end{figure}

Both the transmitter and receiver architectures consist of multiple layers as shown in Fig. \ref{f: arch}. The transmitter passes an initialization waveform $\mathbf{x} $ through the function $f_{\boldsymbol{\theta}_T}(\cdot)$ with the trainable parameter set $\boldsymbol{\theta}_T$. The task of the transmitter is to generate the radar waveform $\mathbf{y}$, which adapts to the actual statistics of the environment so as to improve target detection performance. The transmitted waveform $\mathbf{y}$ has to ensure the waveform power constraint $\mathbf{y}^H \mathbf{y}=1$. For this purpose, a normalization layer is added at the last layer of the transmitter architecture. The receiver passes the received signal $\mathbf{z}$ through the function $f_{\boldsymbol{\theta}_R}(\cdot)$ with the trainable parameter set $\boldsymbol{\theta}_R$. The task of the receiver is to generate a scalar $p\in (0,1)$ that approximates the posterior probability of the presence of a target conditioned on the received signal $\mathbf{z}$ and the receiver parameter set $\boldsymbol{\theta}_R$. To this end, the last layer of the function $f_{\boldsymbol{\theta}_R}(\cdot)$ is selected as a sigmoid function as in logistic regression. The presence and absence of the target is determined based on the output of the receiver and a given threshold, as further discussed in the next section.

\section{Experiments}
In this section, we first introduce the target and clutter models used in the experiments, and then we present numerical results with the aim of assessing the detection performance of the end-to-end learning of the radar system.
\subsection{Target and Clutter Models}
We consider a radar system with a single transmitter, a stationary target, and a single receiver as shown in Fig. \ref{f:end_to_end}. The system aims to detect the presence of the target in a clutter field. The transmitter radar waveform $\mathbf{y}=[y_1, \cdots, y_K]^T$ is composed of $K$ modulated chips with deterministic complex amplitudes $\{y_k \}_{k=1}^K$. If the target is present, i.e., if $m=1$, after chip matched filtering and sampling, the channel $p_m(\mathbf{z}|\mathbf{y})$ outputs the $K \times 1$ discrete-time signal
\begin{equation}
	\mathbf{z}=\mathbf{s}+\mathbf{c}+\mathbf{n}, \label{eq:received signal}
\end{equation}
where $\mathbf{s}=\alpha\mathbf{y}$ denotes the target response with $\alpha$ being the target complex gain, which accounts for target backscattering and channel propagation effects; $\mathbf{c}=\sum_{\substack{ k=-K+1 \\ k\neq0}}^{K-1} \gamma_{k}\mathbf{J}_k\mathbf{y}$ denotes the clutter components, which is the superposition of returns from adjacent range cells, with $\gamma_{k}$ and $\mathbf{J}_k$ being the random clutter scattering coefficient and the shifting matrix associated with the $k$th range cell respectively;
and $\mathbf{n}$ represents the signal-independent interference, which includes the contribution of thermal noise as well as interfering signals due to possible hostile jammers. The shifting matrix $\mathbf{J}_k$ is given by $\big[\mathbf{J}_k  \big]_{i,j}=1$ if $i-j=k$, and $\big[\mathbf{J}_k  \big]_{i,j}=0$ otherwise, with $(i,j)\in \{1,\cdots,K\}^2$. The signal-independent interference $\mathbf{n}\sim \mathcal{CN}(\mathbf{0},\boldsymbol{\Omega}_{n}) $ is assumed to be correlated with correlation matrix $\big[\boldsymbol{\Omega }_n\big]_{i,j}=\sigma _{n}^{2}\rho^{|i-j|}$, where $\sigma_n^2$ is the signal-independent interference power and $\rho$ is the one-lag correlation coefficient. If the target is not present, i.e., if $m=0$, the channel $p_m(\mathbf{z}|\mathbf{y})$ outputs $\mathbf{z}=\mathbf{c}+\mathbf{n}$.

The target is assumed to obey a Swerling Type I model, hence the complex coefficient of target return $\alpha$ is fixed during the observation interval, and has a Rayleigh envelope $\alpha\sim \mathcal{CN}(0,\sigma_{\alpha}^2)$. Clutter components from different range cells are assumed to be uncorrelated, and hence the correlation matrix of the clutter is given by
\begin{equation}
	\boldsymbol{\Omega}_{c}=E[\mathbf{c}\mathbf{c}^H]=\sum_{\substack{ k=-K+1 \\ k\neq0}}^{K-1}\sigma_{c,k}^2 \mathbf{J}_k \mathbf{y} \mathbf{y}^H \mathbf{J}_k^H,
\end{equation}
where $\sigma_{c,k}^2=\text{E}\{|\gamma_k|^2 \}$ denotes the power of the scattering coefficient $\gamma_k$. Clutter scattering coefficients $\{ \gamma_k \}_{k=-K+1,k\neq0}^{K-1}$ are distributed according to coherent Weibull distribution with shape parameter $\beta \in [0.25,2]$ \cite{shape}. Note that when $\beta=2$, clutter scattering coefficients are complex Gaussian random variables. 

From (\ref{eq:received signal}), the detection problem leads to the following binary hypothesis test
\begin{equation}
\left\{ \begin{aligned}
&\mathcal{H}_0\text{ }(m=0):\mathbf{z}=\mathbf{c}+\mathbf{n} \\
&\mathcal{H}_1\text{ }(m=1):\mathbf{z}=\mathbf{s}+\mathbf{c}+\mathbf{n}, \end{aligned} \right.  \label{eq:binary hypo}
\end{equation}
where $\mathcal{H}_0$ is the hypothesis that there is no target $(m=0)$, and $\mathcal{H}_1$ is the hypothesis that a target is present $(m=1)$.

\subsection{Transmitter and Receiver Architectures}
Following Section III, the first layer in the transmitter of Fig. \ref{f: arch} is complex-to-real (C2R) layer, which converts an initialization complex waveform $\mathbf{x}\in \mathbb{C}^K$ into a real one of $2K$ real numbers by separating real and imaginary parts. The transmitter is implemented as a feedforward neural network with $2K$ inputs, $2K$ hidden neurons, and $2K$ outputs. The activation function of the processing neurons is the hyperbolic tangent. The output of the neural network is given by $2K$ reals, which are subsequently converted into a complex vector 
$\mathbf{u}\in \mathbb{C}^K$ through a real-to-complex (R2C) transformation merging two successive real numbers into a complex one. Finally, we obtain the transmitted radar waveform $\mathbf{y}\in \mathbb{C}^K$ through the normalization layer.

The observation from the radar channel is a complex vector $\mathbf{z}\in \mathbb{C}^K$. The C2R layer is adopted at the receiver to convert $\mathbf{z}$ into a real vector. The receiver is implemented as a feedforward neural network with $2K$ inputs, $M$ hidden neurons, and 1 output. The sigmoid function is adopted as the activation function for all neurons at the receiver.
\subsection{Results}
We adopt a stepped frequency waveform \cite{Richards 2005} of length $K=8$ complex-valued chips as the initialization waveform. 
The transmitter and receiver are implemented as feedforward neural networks with parameters $K=8$ and $M=10$, respectively. The training set consists of $4 \times 10^5$ sample vectors equally divided between the $\mathcal{H}_0$ and $\mathcal{H}_1$ hypotheses. For testing, the number of sample vectors is $5 \times 10^6$, equally divided between the $\mathcal{H}_0$ and $\mathcal{H}_1$ hypotheses. We adopt the Gaussian policy described in Section II-B with parameter $\sigma^2=0.3$. The numbers of samples used to estimate the losses in (\ref{eq:CE rx loss}) and (\ref{eq:CE tx loss}) are $Q_R=5\times 10^4$ and $Q_T=4\times 10^5$, respectively. We set the variance of the target complex gain as $\sigma_{\alpha}^2=50$; the power of clutter scattering coefficients as $ \{\sigma_{c,k}^2 \}_{k=-K+1,k\neq0}^{K-1} =1/7$; the signal-independent power as $\sigma_n^2=1$; and the one-lag correlation coefficient as $\rho=0.4$. We denote $\beta_{\text{train}}$ and $\beta_{\text{test}}$ as the shape parameters of the clutter distribution applied in training and test stage, respectively. 

As performance measures, we adopt the standard probability of detection $\text{P}_\text{d}$ and probability of false alarm $\text{P}_{\text{fa}}$. The receiving operating characteristic (ROC) curves are obtained via Monte Carlo simulations by varying the threshold applied to the output of the receiver.

\begin{figure}[H]
	\centering
	\vspace*{-2ex}
	\includegraphics[width=0.6\linewidth]{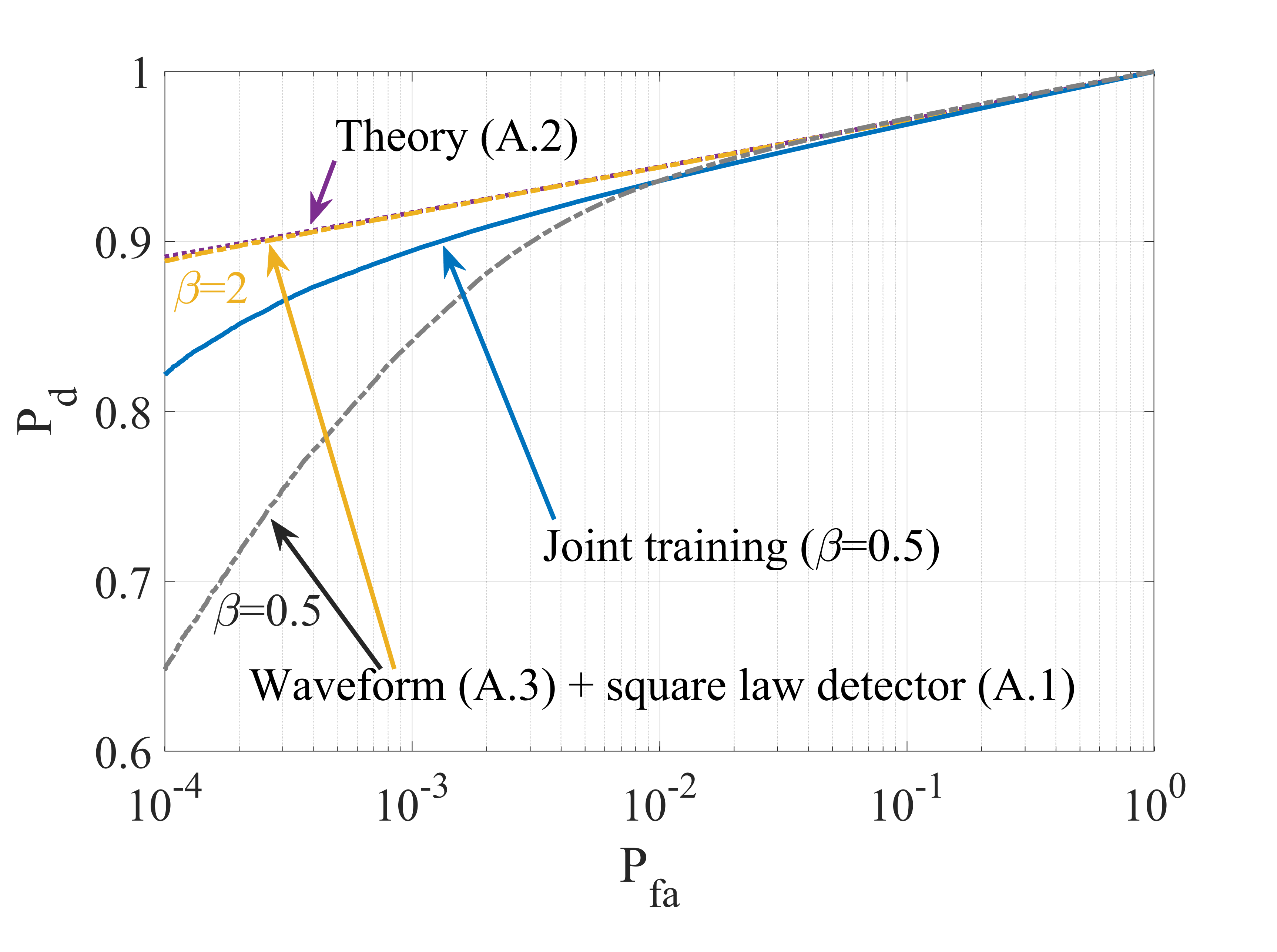}
	\vspace*{-3ex}
	\caption{ROC curves for Gaussian/non-Gaussian clutter with the same shape value $\beta_{\text{train}}=\beta_{\text{test}}=\beta$ in training and test phases.}
	\label{f:fig1}
\end{figure}

Fig. \ref{f:fig1} compares the ROC curves for Weibull clutter with different values of the shape parameter $\beta$ that is assumed to be equal for both train and test phases, i.e., $\beta_{\text{test}}=\beta_{\text{train}}=\beta$. When $\beta=2$, i.e., for Gaussian clutter, the optimal waveform and detector are available and given by (\ref{eq:opt y}) and by the square law detector (\ref{eq: test}) in Appendix A. For this case, the ROC curve can also be computed in closed form as $\text{P}_{\text{d}}=\text{P}_{\text{fa}}^{{1}/\big[ {1+\sigma_{\alpha}^2\mathbf{y}^H \big( \boldsymbol{\Omega}_{c}+\boldsymbol{\Omega}_{n} \big)^{-1} \mathbf{y} \big]}}$. In contrast, when the clutter is non-Gaussian, i.e., $\beta\neq2$, the optimal detector is not known, while the optimal waveform does not depends on clutter distribution. As observed in the figure, for a shape parameter $\beta=0.5$, the proposed end-to-end learning approach for the joint design of waveform and detector provides significant gains over waveform (\ref{eq:opt y}) and square law detector (\ref{eq: test}).

\begin{figure}[H]
	\vspace{-3ex}
	\centering
	\includegraphics[width=0.6\linewidth]{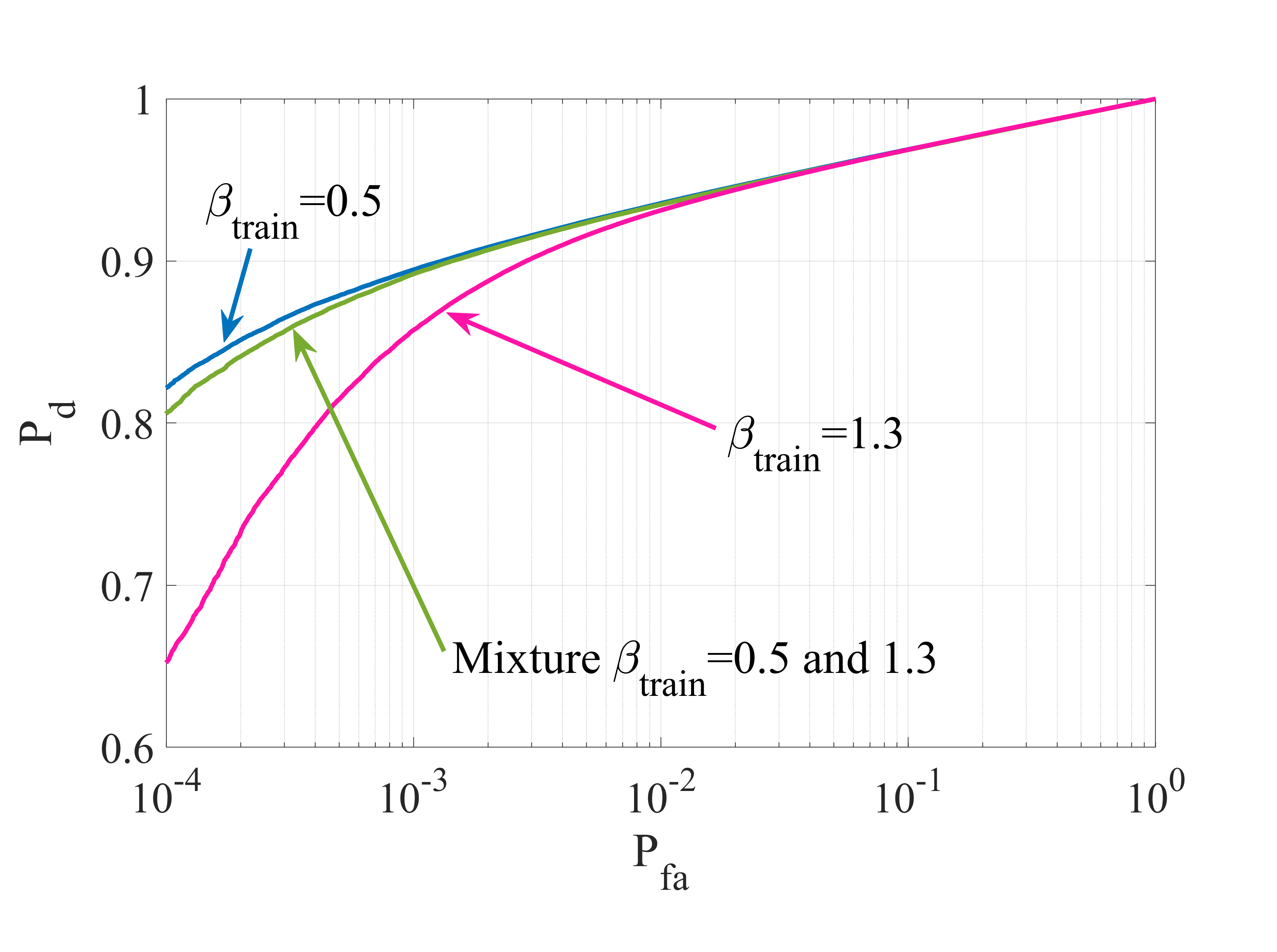}
	\vspace{-4ex}
	\caption{ROC curves for non-Gaussian clutter with joint training under different clutter statistics between testing and training when $\beta_{\text{test}}=0.5$.}
	\label{f:fig2}
\end{figure}

Fig. \ref{f:fig2} illustrates the robustness of the trained radar system to changes in the clutter statistics. Instead of training by assuming a single shape parameter, we propose here to robustify the system by drawing samples in (\ref{eq: est CE rx loss}) and (\ref{eq:tx loss grad}) from a mixture of distributions, while testing for one value $\beta_{\text{train}}$. As shown in the figure, the end-to-end leaning radar system trained by mixing clutter samples with $\beta_{\text{train}}=$ 0.5 and 1.3 outperforms the system trained by assuming $\beta_{\text{train}}=1.3$ when tested with $\beta_{\text{test}}=0.5$.

\begin{figure}[H]
	\centering
	\vspace{-2ex}
	\includegraphics[width=0.6\linewidth]{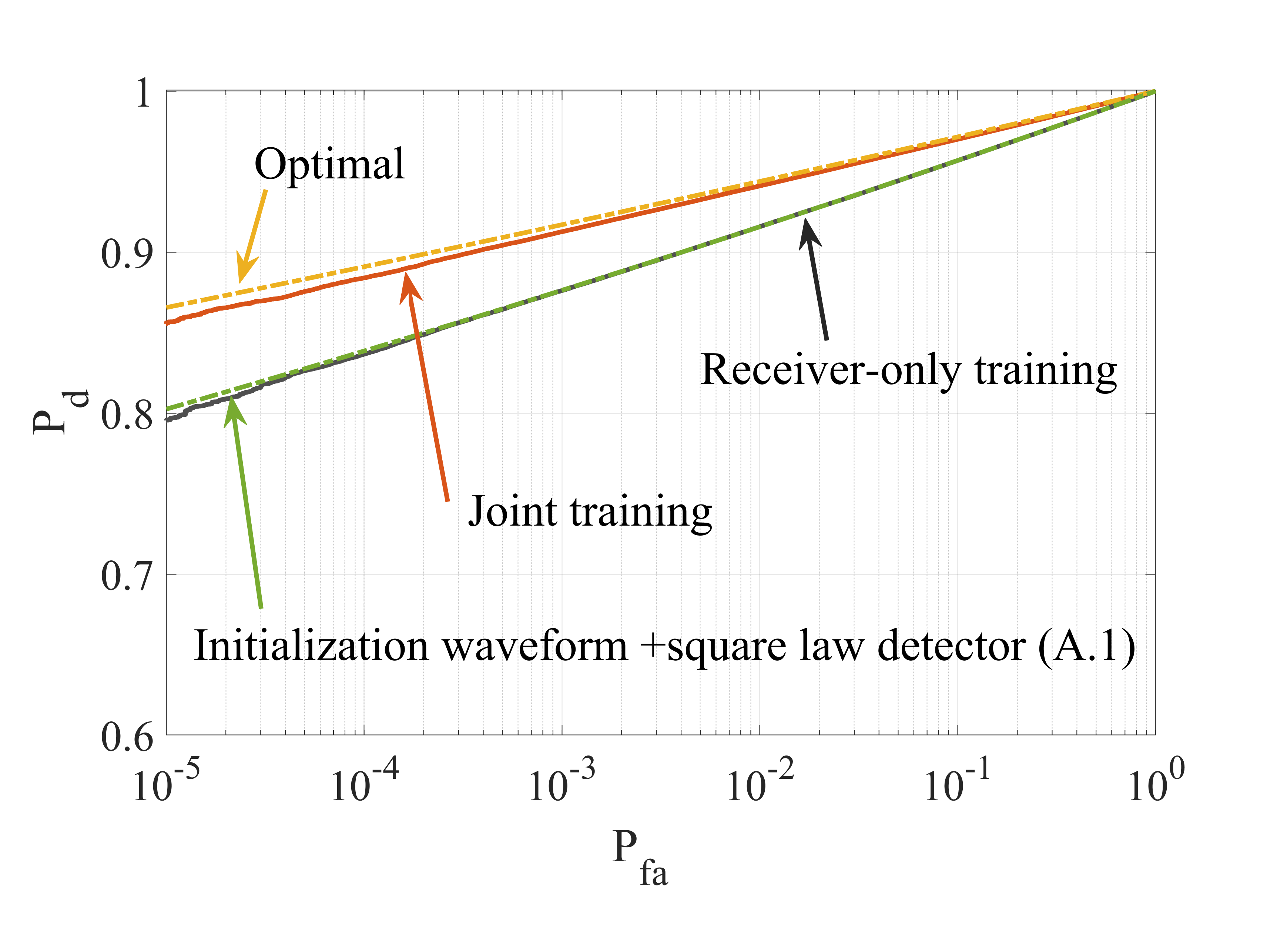}
	\vspace{-4ex}
	\caption{ROC curves with and without training for Gaussian clutter ($\beta_{\text{train}}=\beta_{\text{test}}=2$).}
	\vspace{-1ex}
	\label{f:fig3}
\end{figure}

Finally, Fig. \ref{f:fig3} compares the ROC curves with joint training and with only receiver-side training for $\beta_{\text{train}}=\beta_{\text{test}}=2$. Joint training is seen to result in a significant improvement of the ROC as compared to training only the receiver. Moreover, joint training provides a comparable detection performance with the optimal waveform (\ref{eq:opt y}) and square law detector (\ref{eq: test}).

\section{Conclusions}
In this paper, we have formulated the radar design problem as the end-to-end learning of waveform generation and detection. We have developed a joint training algorithm that iterates between supervised training of the receiver and reinforcement learning-based training of the transmitter. We have also proposed to robustify the detection performance by training the system with mixed clutter statistics. Numerical results have shown that the proposed end-to-end leaning approach is beneficial under non-Gaussian clutter.

\numberwithin{equation}{section} 
\appendices
\section{}
When the shape parameter of Weibull distribution is $\beta=2$, i.e., under a Gaussian clutter, the optimal radar waveform and detector for (\ref{eq:binary hypo}) are known and reviewed here. The optimal detector in the Neyman-Pearson sense is the square law detector \cite{Richards 2005}, which is given by the test
\begin{equation}
\bigg|\mathbf{z}^H \big( \boldsymbol{\Omega}_{c} + \boldsymbol{\Omega}_{n} \big)^{-1}\mathbf{y} \bigg|^2 \mathop{\gtrless}%
_{\mathcal{H}_0}^{\mathcal{H}_1}\epsilon, \label{eq: test}
\end{equation}
where $\epsilon$ is the detection threshold. Also, an analytical expression of the detection probability $\text{P}_{\text{d}}$ as a function of the false alarm probability $\text{P}_{\text{fa}}$ is available, and is given as
\begin{equation}
\text{P}_{\text{d}}=\text{P}_{\text{fa}}^{{1}/\big[ {1+\sigma_{\alpha}^2\mathbf{y}^H \big( \boldsymbol{\Omega}_{c}+\boldsymbol{\Omega}_{n} \big)^{-1} \mathbf{y} \big]}}. \label{eq: pd vs pfa}
\end{equation}
From (\ref{eq: pd vs pfa}), the probability $\text{P}_{\text{d}}$ is an monotonically increasing function of $\mathbf{y}^H \big( \boldsymbol{\Omega}_{c}+\boldsymbol{\Omega}_{n} \big)^{-1} \mathbf{y} $. Thus, the optimal radar waveform for target detection can be obtained by solving the following problem
\begin{equation}
\begin{aligned}
& \underset{\mathbf{y}}{\text{max}}&&
\mathbf{y}^H \big( \boldsymbol{\Omega}_{c}+\boldsymbol{\Omega}_{n} \big)^{-1} \mathbf{y}  \\
& \text{s.t.} && \mathbf{y}^H \mathbf{y}=1.
\end{aligned} \label{eq:opt y}
\end{equation}
The optimal solution to the optimization problem (\ref{eq:opt y}) could be obtained via the sequential optimization algorithm in \cite{adaptive waveform}.

\end{document}